\documentclass[aps,prl,twocolumn,floats,floatfix,english]{revtex4}
\usepackage[T1]{fontenc}
\usepackage[latin1]{inputenc}
\usepackage{amsmath}
\usepackage{setspace}
\usepackage{epsfig}

\makeatletter

\providecommand{\LyX}{L\kern-.1667em\lower.25em\hbox{Y}\kern-.125emX\@}

\usepackage{babel}
\makeatother
\begin{document}
\title{Intelligent Minority Game with genetic-crossover
strategies}

\author{Marko Sysi-Aho}
\email[]{msysiaho@lce.hut.fi}
\author{Anirban Chakraborti}
\email[]{anirban@lce.hut.fi}
\author{Kimmo Kaski}
\email[]{Kimmo.Kaski@hut.fi}

\affiliation{Laboratory of Computational Engineering, Helsinki
University of Technology, \\
P. O. Box 9203, FIN-02015 HUT, Finland.}

\begin{abstract}

We develop a game theoretical model of $N$ heterogeneous
interacting agents called the intelligent minority game.
The ``intelligent'' agents play the basic minority game and depending
on their performances, generate new strategies using the one-point genetic 
crossover mechanism. The performances change dramatically and the
game moves rapidly to an efficient state (fluctuations in the number of 
agents performing a particular action, characterized by $\sigma^2$, reaches a low value). There is no ``phase 
transition'' when we vary $\sigma^2/N$ with $2^M/N$, where $M$ is the ``memory''
of an agent.

\end{abstract}

\maketitle


The dynamics of interacting agents competing for scarce resources
are believed to underlie the behaviour of complex systems in natural \cite{parisi,huberman,nowak} 
and social \cite{lux,arthur} sciences.
The agents have to be the best in order to survive-- similar to the
idea of {}``survival of the fittest'' in biology. In studies of market
behaviour, tools of statistical physics have been combined with
theories of economics
\cite{Econophy1,Econophy2,Econophy3,Econophy4}, like game theory,
which deals with decision making of a number of rational
opponents under conditions of conflict and competition
\cite{game,challet1,challet2,cavagna,riolo,lamper}. 

In this letter, we present a game theoretical model of a large
number of heterogeneous interacting agents called the intelligent
minority game, based on the minority game \cite{challet1}. This provides an alternative
to the representative approach of microeconomics, where one has a
theory with a single (representative) agent, based on the assumption
that all the agents are identical \cite{micro}. The minority game model consists of agents having a
finite number of strategies and finite amount of public information,
interacting through a global quantity (whose value is fixed by all
the agents) representing a market mechanism. In the original
model the agents choose their strategy through a simple adaptive dynamics
based on \emph{inductive reasoning} \cite{arthur}. Here, we introduce
the fact that the agents are \emph{intelligent} and in order to be
the best or survive in the market, modify their strategies
periodically depending on their performances.
For modifying the strategies, we choose the mechanism of
\emph{one-point genetic crossover}, following the ideas of genetic algorithms
in computer science and operations research. In fact, these algorithms were inspired
by the processes observed in natural evolution
\cite{holland,goldberg,lawrence} and it turned out that they solve some
extremely complicated problems without knowledge of the decoded
world. In nature, one-point crossover occurs when two parents
exchange parts of their corresponding chromosomes after a selected
point, creating offsprings \cite{lawrence}.


The basic minority game consists of an odd number of agents $N$
who can perform only two actions, at a given time $t$, and an agent
wins the game if it is one of the members of the minority group.
The two actions, such as {}``buying'' or {}``selling'' commodities, are
denoted here by $0$ or $1$. Further, it is assumed that all the agents have access to finite amount
of public information, which is a common bit-string {}``memory''
of the $M$ most recent outcomes. Thus the agents are said to exhibit
{}``bounded rationality'' \cite{arthur}. For example, in case of
memory $M=2$ there are $P=2^{M}=4$ possible {}``history''
bit strings: $00$, $01$, $10$ and $11$. A {}``strategy'' consists
of a response, i.e., $0$ or $1$, to each possible history bit strings;
therefore, there are $G=2^{P}=2^{2^{M}}=16$ possible strategies which
constitute the {}``total strategy space''. In our study, we use the reduced
strategy space by picking only the uncorrelated strategies (which
have Hamming distance $d_{H}=1/2$) \cite{challet3}. At the beginning
of the game, each agent randomly picks $k$ strategies, and after a
game, assigns one {}``virtual'' point to the strategies which would
have predicted the correct outcome; the best strategy is the one
which has the highest virtual point. The performance of
the player is measured by the number of times the player wins, and
the strategy, which the player uses to win, gets a {}``real''
point. We also keep a record of the number of agents who have chosen a particular
action, say, {}``selling'' denoted by $1,$ $N_{1}(t)$ as a function
of time. The fluctuations in the behaviour of $N_{1}(t)$
indicate the total utility of the system. For example, we may have a situation
where only one player is in the minority and thus wins, and all the
other players lose. The other extreme case is when $(N-1)/2$ players are in the
minority and $(N+1)/2$ players lose. The total utility of the system
is highest for the latter case as the total number of the
agents who win is maximum. Therefore, the system is more
efficient when there are smaller fluctuations around the mean than
when the fluctuations are larger. The fluctuations can be characterized by the variance $\sigma^2$
so that smaller values of $\sigma^2$ would correspond to a more efficient state.

In our model, the players of the basic minority game are
assumed to be intelligent and modify their
strategies after every time-interval $\tau$ depending on their performances.
If they find that they are among the fraction $n$
(where $0<n<1$) of the worst performing players, they modify any two
of their strategies chosen randomly from the pool of $k$ strategies and use one of the new
strategies generated.
The mechanism by which they modify their strategies is that of one-point
genetic crossover illustrated schematically in Figure 1. The
strategies $s_{i}$ and $s_{j}$ act as the parents
and by choosing the breaking point in them randomly, and performing
one-point genetic crossover, the children $s_{k}$ and $s_{l}$ are
produced. We should note that the strategies are changed by the agents
themselves and even though the strategy space evolves, it is still
of the same size and dimension; thus considerably different from
earlier attempts \cite{challet1,li1,li2}. 

\begin{figure}
\epsfig{file=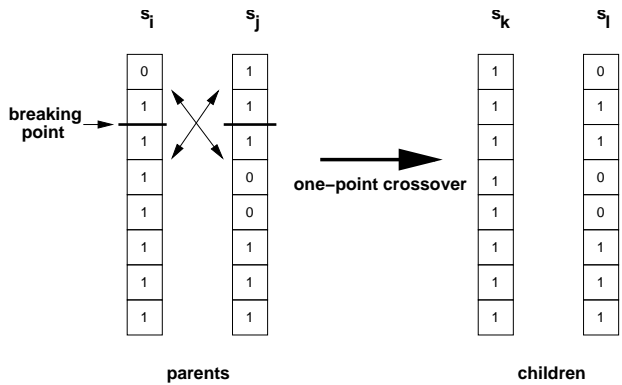,width=3.2in }
\caption{
Schematic diagram to show the mechanism
of one-point genetic crossover to produce new strategies. The strategies
$s_{i}$ and $s_{j}$ are the parents. We choose the breaking point
randomly and through this one-point genetic crossover, the children
$s_{k}$ and $s_{l}$ are produced.
}
\label{fig1}
\end{figure}

\begin{figure}
\epsfig{file=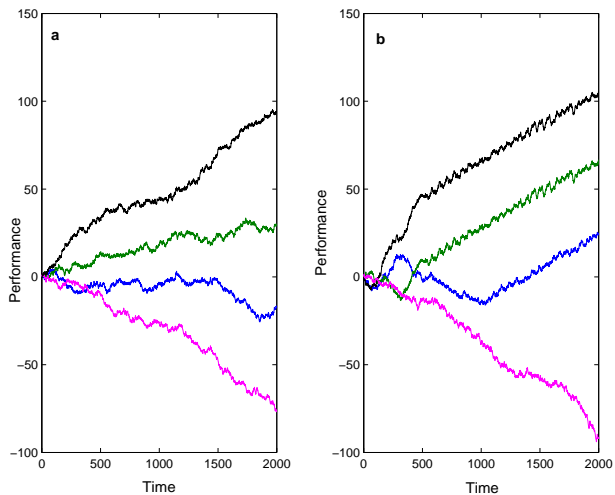,width=3.2in }
\caption{
Plots of the performances of the best
player (black), the worst player (magenta) and two randomly selected
players (green and blue) in (a) the basic minority game, where $N=1001$, $M=5$, $k=10$
and $t=1999$, and (b) in the intelligent minority game, where
$N=1001$, $M=5$, $k=10$, $t=1999$, $n=0.3$ and $\tau =100$.
}
\label{fig2}
\end{figure}

In Figure 2,
the performances of the players in our model are compared with those
in the basic minority game. We have scaled the performances of all the
players such that the mean is zero for easy comparison of the
success of the agents in each case. We find that there are significant
differences in the performances of the players. The performance of
a player in the basic minority game does not change drastically in the course of the
game as shown in Figure 2 (a). However, in our model, the
performances of the players
may change dramatically even after initial downfalls, and agents often
do better after they have produced new strategies with the one-point
genetic crossovers, as illustrated in Figure 2 (b). 

\begin{figure}
\epsfig{file=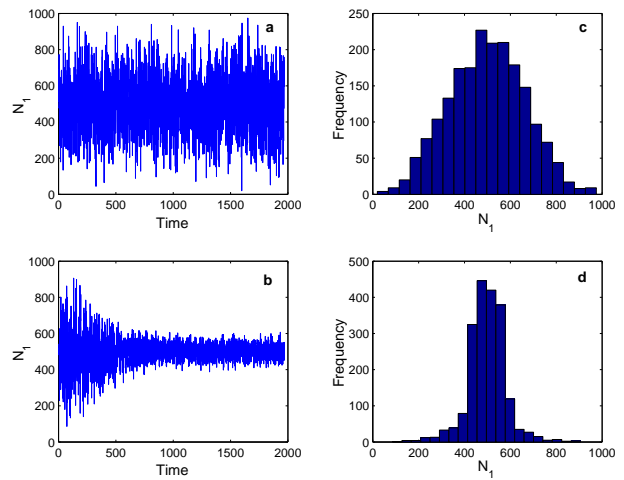,width=3.2in }
\caption{
Plots of the (a) time-variation of $N_{1}$ for the basic minority game  (b) time-variation of 
$N_{1}$ for the intelligent minority game (c) histogram of $N_{1}$ for the basic minority game
 and (d) histogram of $N_{1}$ for the intelligent minority game.
The simulations for the basic minority game have been made with $N=1001$, $M=5$,
$k=10$ and $t=1999$ and for the intelligent minority game with $N=1001$, $M=5$, $k=10$,
$t=1999$, $n=0.3$ and $\tau =100$.
}
\label{fig3}
\end{figure}

In order to study the efficiency of the game, 
we plot the time-variation of $N_{1}$
for the basic minority game in comparison to our model in Figures 3 (a) and (b). Also
the histograms of $N_{1}$ for the basic minority game and our model are plotted in
Figures 3 (c) and (d). Clearly evident from these figures is the fact
that when we allow one-point genetic crossovers in strategies, the
system moves toward a more efficient state since the fluctuations
in $N_{1}$ decreases and the histogram of $N_{1}$ becomes narrower
and sharper. We have also studied the effect of increasing the fraction
of players $n$ on the distributions of the number of switches and
the number of genetic crossovers the players make. The results in
Figure 4
illustrate the fact that as $n$ increases, more players have to
make large number of switches and crossovers in order to be the
best.

\begin{figure}
\epsfig{file=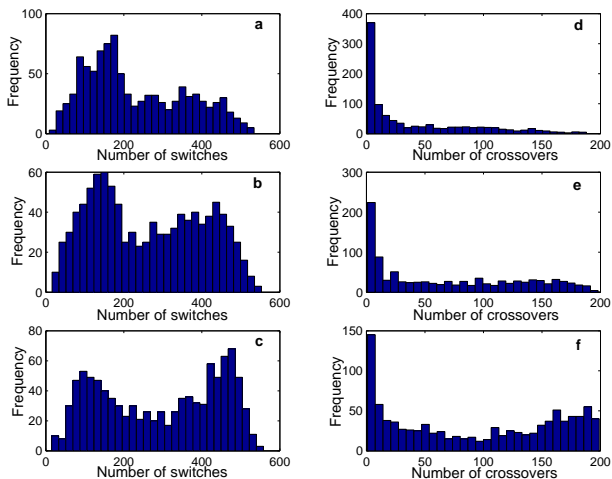,width=3.2in }
\caption{
The histograms of the number of switches
the players make in the intelligent minority game for (a) $n=0.3$ (b) $n=0.4$ (c) $n=0.5$,
and the histograms of the number of genetic crossovers the players
make in the intelligent minority game for (d) $n=0.3$ (e) $n=0.4$ and (f) $n=0.5$. The
simulations have been made with $N=1001$, $M=4$, $k=10$, $t=1999$
and $\tau =10$.
}
\label{fig4}
\end{figure}

Furthermore, we calculate the variance $\sigma^2$ of $N_{1}$.
The variation of $\sigma^2/N$ against the parameter $2^M/N$ for the basic minority game, have been studied in details in refs. \cite{challet2,challet3,li1,li2}.
We show the variation of $\sigma^2/N$ with the parameter $2^M/N$
for $k=2$ in Figure 5 (a) for both the games, by varying $M$ and $N$. 
Also, we plot the quantity $\sigma^2/N$ against $M$ (varied 
from 2 to 12) for $N=1001$ players and different values of $k$, in
Figure 5 (b). For $k=2$, the quantity $\sigma^2/N$ is minimum
in the basic minority game when $2^M/N \approx 0.5$ and there is a ``phase transition'' at this value \cite{challet2,challet3,li1,li2}.
As we increase the value of $k$ the efficiency decreases and this transition finally smoothens out.
However, in the intelligent minority game, we find no such phase
transition for any combinations of $k$, $M$ and $N$, we have
studied. We found that as  the value of $k$ is increased, the efficiency decreases, but at a rate much smaller than in the basic minority game. For both games, the values of $\sigma^2/N$ seem to converge 
towards a common value for large values of $M$. If we compare the two games, we
find that for large $k$ values and moderate values of $M$, 
the differences in $\sigma^2/N$ is very large.

\begin{figure}
\epsfig{file=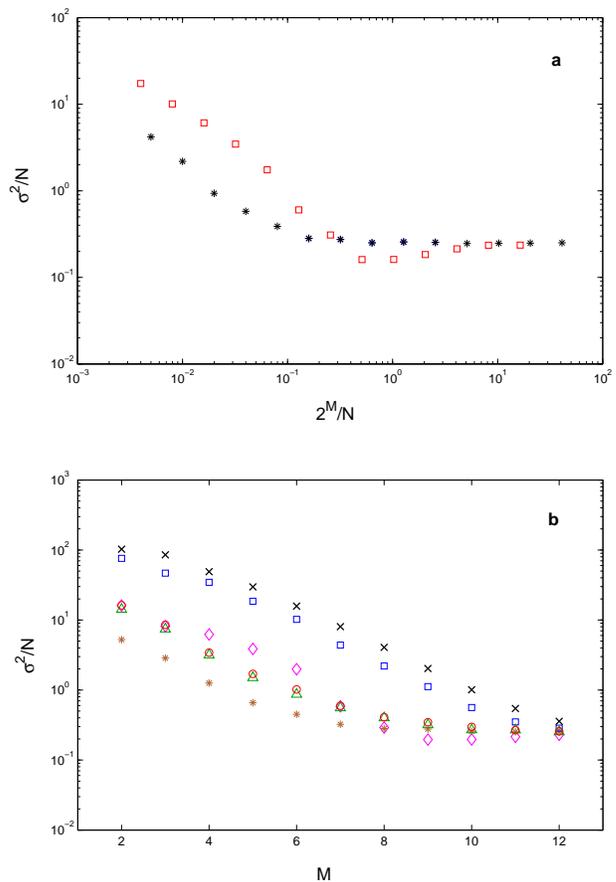,width=3.2in }
\caption{
(a) The plot of $\sigma^2/N$ against the parameter $2^M/N$ for $k=2$, by
varying $M$ from $2$ to $11$ and $N$ from $25$ to $1001$ for the
basic minority game (red squares) and the intelligent minority game
(black asterisk marks). The
simulations were made for $t=5000$ and ten different samples in
each case. The parameter values chosen for the intelligent minority
game were $\tau=10$ and $n=50$.
(b) The plot of $\sigma^2/N$ against $M$ for different values of
$k$ for the basic minority game and the intelligent 
minority game. For the basic minority game, we have studied the
cases of $k=2$ (magenta diamonds), $k=6$ 
(blue squares) and $k=10$ (black cross marks). For the intelligent minority game, we have studied 
the cases of $k=2$ (brown asterisk marks), $k=6$ (green triangles) and $k=10$ (red circles). The simulations
for the basic minority game have been made with $N=1001$ and  $t=5000$, and for the intelligent 
minority game have been made with $N=1001$, $t=5000$, $n=50$ and
$\tau =10$, and for five different samples in each case.
}
\label{fig5}
\end{figure}

We have observed that
in our model, the worst players were often those who switched strategies
most frequently while the best players were those who made the least number
of switches after finding a good strategy. Further, we found that
the players who do not make any genetic crossovers are unable to compete
with those who make genetic crossovers, and their performances were
found to fluctuate around the zero mean. Moreover, it was found
that as the
crossover time-interval $\tau $ is increased, the time for the
system to reach an efficient state is longer \cite{Marko}.


One advantage of our model is clearly that
the dimensionality of the strategy space as well as the number of
elements in the strategy space remain the same. It is also appealing
that starting from a small number of strategies, many ``good'' strategies
can be generated by the players in the course of the game.
Even though the players may not have performed well initially, they
often did better when they used new strategies generated by the one-point
genetic crossovers. Finally, it should be pointed out that even in
the framework of genetic algorithms, there are various ways to
generate new strategies. One possibility is that we make a
one-point genetic crossover between the two worst strategies and
replace the parents by the children. Another possibility is to make
``hybridized genetic crossover''  where we make a one-point genetic
crossover between the two best strategies, replace the
worst two strategies with the children and retain
the parents as well. We defer these modifications and interesting results for
a future communication \cite{Marko}. 
\vskip 0.25in

\begin{acknowledgments}
This research was partially supported by the Academy of
Finland, Research Centre for Computational Science and Engineering,
project no. 44897 (Finnish Centre of Excellence Programme 2000-2005).
\end{acknowledgments}


\noindent \vskip 0.25in

\begin{thebibliography}{10}
\bibitem{parisi}G. Parisi, \textit{Physica A} \textbf{263}, 557 (1999).
\bibitem{huberman}B. A. Huberman, P. L. T. Pirolli, J. E. Pitkow and R. M. Lukose, \textit{Science}
\textbf{280}, 95 (1998).
\bibitem{nowak}M. Nowak and R. May, \textit{Nature} \textbf{359}, 826 (1992).
\bibitem{lux}T. Lux and M. Marchesi, \textit{Nature} \textbf{397}, 498 (1999).
\bibitem{arthur}W. B. Arthur, \textit{Am. Econ. Rev.} \textbf{84}, 406 (1994).
\bibitem{Econophy1}R. N. Mantegna and H. E. Stanley, \textit{An Introduction to Econophysics},
Cambridge University Press, Cambridge (2000).
\bibitem{Econophy2}J.-P. Bouchaud and M. Potters, \emph{Theory of Financial Risk}, Cambridge
University Press, Cambridge (2000).
\bibitem{Econophy3}S. Moss de Oliveira, P. M. C. de Oliveira and D. Stauffer, \textit{Evolution,
Money, War and Computers}, B. G. Teubner, Stuttgart-Leipzig (1999).
\bibitem{Econophy4}J. D. Farmer, \emph{preprint available on adap-org/9912002} (1999).
\bibitem{game}R. Myerson, \emph{Game Theory: Analysis of Conflict}
(Harvard University Press, Cambridge, Massachusetts, 1991).
\bibitem{challet1}D. Challet and Y.-C. Zhang, \textit{Physica A} \textbf{246}, 407 (1997).
\bibitem{challet2}D. Challet, M. Marsili and R. Zecchina, \textit{Phys. Rev. Lett.} \textbf{84}, 
1824 (2000).
\bibitem{riolo}R. Savit, R. Manuca and R. Riolo, \textit{Phys. Rev. Lett.} \textbf{82}, 2203 (1999).
\bibitem{cavagna}A. Cavagna, J. P. Garrahan, I. Giardina and D. Sherrington, \textit{Phys. Rev. Lett.} \textbf{83}, 4429 (1999).
\bibitem{lamper}D. Lamper, S. D. Howison and N. F. Johnson, \textit{Phys. Rev. Lett.}
\textbf{88}, 17902 (2002).
\bibitem{micro}A. Mas-Colell, M. D. Whinston and J. R. Green, \textit{Microeconomic
Theory}, Oxford University Press, New York (1995).
\bibitem{holland}J. H. Holland, \textit{Adaptation in Natural and Artificial Systems},
University of Michigan Press, Ann Arbor (1975).
\bibitem{goldberg}D. E. Goldberg, \textit{Genetic Algorithms in Search, Optimization
and Machine Learning}, Addison-Wesley, Reading, Massachusetts (1989).
\bibitem{lawrence}D. Lawrence (Ed.), \textit{Handbook of Genetic Algorithms}, Van Nostrand
Reinhold, New York (1991).
\bibitem{challet3}D. Challet and Y.-C. Zhang, \textit{Physica A} \textbf{256}, 514 (1998).
\bibitem{li1}Y. Li, R. Riolo and R. Savit, \emph{Physica A} \textbf{276}, 234 (2000).
\bibitem{li2}Y. Li, R. Riolo and R. Savit, \emph{Physica A} \textbf{276}, 265 (2000).
\bibitem{Marko}M. Sysi-Aho, A. Chakraborti and K. Kaski, in preparation (2002).
\end{thebibliography}
\end{document}